\documentclass[prl,twocolumn,superscriptaddress,showpacs]{revtex4}
\usepackage{graphicx}%
\usepackage{dcolumn}
\usepackage{bm}
\begin{document} 

\title{Bonding Configurations and Collective Patterns of Ge Atoms Adsorbed
       on Si(111)-$7\times7$}

\author{Y.~L. Wang}
\affiliation{Beijing National Laboratory for Condensed Matter Physics,
             Institute of Physics, Chinese Academy of Sciences, Beijing
             100080, China} 
\author{H.~-J. Gao}
\altaffiliation{Corresponding authors. E-mail: hjgao@aphy.iphy.ac.cn
                (H.-J.G.); pantelides@vanderbilt.edu (S.T.P.).}
\affiliation{Beijing National Laboratory for Condensed Matter Physics,
             Institute of Physics, Chinese Academy of Sciences, Beijing
             100080, China}
\author{H.~M. Guo}
\affiliation{Beijing National Laboratory for Condensed Matter Physics,   
             Institute of Physics, Chinese Academy of Sciences, Beijing
             100080, China}
\author{Sanwu Wang}
\affiliation{Department of Physics and Astronomy, Vanderbilt University,
             Nashville, Tennessee 37235, USA}
\author{Sokrates T. Pantelides}
\altaffiliation{Corresponding authors. E-mail: hjgao@aphy.iphy.ac.cn
                (H.-J.G.); pantelides@vanderbilt.edu (S.T.P.).}
\affiliation{Department of Physics and Astronomy, Vanderbilt University,
             Nashville, Tennessee 37235, USA}
\affiliation{Condensed Matter Sciences Division, Oak Ridge National
             Laboratory, Oak Ridge, Tennessee 37831, USA\\}

\begin{abstract}

We report scanning tunneling microscopy observations of Ge deposited on
the Si(111)-$7\times7$ surface for a sequence of sub-monolayer coverages.
We demonstrate that Ge atoms replace so-called Si adatoms. Initially, the
replacements are random, but distinct patterns emerge and evolve with
increasing coverage, till small islands begin to form. Corner adatom sites
in the faulted half unit cells are preferred. First-principles density
functional calculations find that adatom substitution competes
energetically with a high-coordination bridge site, but atoms occupying
the latter sites are highly mobile. Thus, the observed structures are
indeed more thermodynamically stable.

\end{abstract}

\pacs{68.43.-h, 68.37.Ef, 73.20.-r, 68.47.Fg}

\maketitle

Semiconductor nanostructures have been mostly fabricated using compound 
semiconductors, benefiting from a continuous variation of one or more 
elements. The Si/Ge system is more limited, but has the advantage that it
is naturally compatible with Si technology. Indeed, Ge is currently
incorporated in Si structures to fabricate strained Si layers with 
enhanced mobility. There is, therefore, renewed activity in Ge-based
nanostructures grown on Si
\cite{Pearsall,Schowalter,Alivisatos,Westphal,Gao1,Lobo,Gao2,Ratto}.
The Si(111)-$7\times7$ surface offers unique potential for the
self-assembly of diverse structures because of the large number of
distinct bonding sites. ``Magic'' Si islands and metal nanoclusters were
recently grown on this surface \cite{Voig,Li,Wu}. Yet, twenty years of
studies have not led to definitive conclusions about the initial bonding
structures of Ge atoms on this surface, impeding further understanding 
and potential control of the growth process.

Sub-monolayer Ge adsorbates on Si(111)-$7\times7$ were investigated using
x-ray standing-wave (XSW) measurements at 300~$^\circ$C by Patel {\it et
al.} in 1985 but it was not possible to determine the precise Ge sites and
the bonding structure \cite{Patel}. Using XSW measurements, Dev {\it et
al.} proposed in 1986 that, at low coverages ($<$~0.5~ML), Ge atoms would
prefer to occupy the on-top sites and to bond directly to the Si adatoms
and restatoms (see Fig. 1 for a schematic of the Si(111)-$7\times7$
surface and pertinent terminology) \cite{Dev}. Reflection electron
microscopy and transmission electron diffraction investigations by
Kajiyama {\it et al.} in 1989 on Ge/Si(111)-$7\times7$ prepared at
640~$^\circ$C found evidence that Ge atoms randomly substituted any Si
atoms at the top layers \cite{Kajiyama}. Then, core-level photoemission 
spectroscopy measurements by Carlisle {\it et al.} in 1994 provided
indirect evidence that there was some preference for Ge to replace the Si
adatoms in the case of annealed Ge/Si(111)-$7\times7$ samples
\cite{Carlisle1}. More recent measurements using near-edge x-ray
absorption spectroscopy and scanning tunneling microscopy (STM) did not
provide conclusive descriptions of Ge bonding sites on the 
Si(111)-$7\times7$ surface \cite{Castrucci,Gao2,Takaoka}. Very few
theoretical calculations have been reported on Ge bonding sites on
Si(111)-$7\times7$. Early work was semiempirical with limited predictive
capabilities, but provided support for the notion that Ge atoms bond
directly to Si restatoms or adatoms \cite{Grodzicki,Stauffer2}. In 1998,
Cho and Kaxiras reported a limited exploration of bonding possibilities
using first-principles calculations and found that the most stable 
adsorption position for Ge on Si(111) is the high-coordination bridge
($B_2$) site, a bonding site that had not been proposed as likely on the 
basis of experimental data \cite{Cho1}.

In this Letter, we report STM observations and first-principles
calculations for the structure of the Ge/Si(111)-$7\times7$ surface at low
Ge coverages. Direct STM observations clearly show that, at low coverages,
Ge atoms reside at the Si adatom sites. Profile measurements rule out the
possibility of co-existence of a Ge atom and a Si adatom underneath it,  
thus revealing that the Ge atoms substitute for the Si adatoms. Initially
(up to 0.02 ML), the occupation of adatoms sites is random with a slight
preference for corner adatoms in the faulted half unit cell (FHUC). As
coverage is increased, the preference for the FHUC corner adatom sites is
enhanced. At 0.08 ML, a distinct triangular pattern of Ge atoms at the
corners of the FHUC is dominant. At a slightly higher coverage (0.1 ML) 
other distinct patterns become more visible and tiny islands start to 
appear. The above observations are complemented with first-principles
calculations. We find that the high-coordination $B_2$ configuration of
adsorbed Ge has roughly the same energy as the configuration in which a Ge
atom replaces a Si adatom, with the latter occupying the lowest-energy 
nearby site. However, we also find the atoms at the $B_2$ sites are highly 
mobile, whereas Ge atoms that replace Si adatoms are very stable against
diffusion.

The experiments were conducted in an ultrahigh-vacuum STM system (Omicron
UHV-STM, Germany) with a base pressure $\sim$5$\times$10$^{-11}$~mbar. The
samples were cut from an antimony-doped {\it n}-type Si(111) wafer
(resistance: $\rho\sim0.03~\Omega\cdot$cm; thickness: $\sim0.5$~mm).
Before it was introduced into the vacuum chamber, the sample was cleaned
by ethanol in an ultrasonic bath and rinsed thoroughly by de-ionized
water. Inside the chamber it was de-gassed for several hours at
$\sim600~^\circ$C. The sample was annealed by direct current heating while
the pressure was kept below 5$\times10^{-10}$~mbar. An annealing cycle
consisted of flashing the sample to $1200~^\circ$C for 20 seconds and
lowering the temperature fast to about $900~^\circ$C and then at a slow
decreasing pace rate of $1-2~^\circ$C/s to room temperature.
Si(111)-$7\times7$ reconstructed surface was finally obtained. Ge
(99.9999\% purity) was deposited onto the as-prepared Si(111)-$7\times7$
surface by resistive evaporation and the substrate temperature was
$150~^\circ$C by irradiation. During evaporation the pressure in the
chamber was lower than 5$\times10^{-10}$~mbar. A typical deposition rate
of $\sim$~0.01 ML/min was routinely achieved. One monolayer is defined as
the atomic density of the unreconstructed Si(111) surface (1 ML =
7.83$\times10^{14}$~atoms/cm$^2$). All the STM images were acquired in a
constant-current mode with an electrochemically etched tungsten tip at
room temperature.

Figure 2 shows STM topographic images of the Si(111)-$7\times7$ surface
with Ge coverages of 0.02 ML, 0.08 ML, and 0.10 ML, respectively. These
images show that the surface lattice retains the original $7\times7$
reconstruction. The dimers and the Si adatoms are visible. The FHUC and
the unfaulted half unit cell (UHUC) of the $7\times7$ reconstruction are
distinguished due to the different contrast [Fig. 2(a)]. The deposited Ge
atoms appear as bright protrusions. Three significant features are present
in the STM images. First, the deposited Ge atoms are clearly resolved as
single atom. Second, the adsorbed Ge atoms reside on the sites that were
occupied by the Si adatoms on Si(111)-$7\times7$. Finally, more Ge atoms
occupy the corner adatom sites in the FHUC than the other adatom sites.
{\it No Ge atoms are found at either the restatom or the high-coordination
surface sites}. Furthermore, profile lines through the bright dots in the   
STM images show that the height difference between the Ge atom and the
original Si adatoms is about 0.2~{\AA}, as shown in Fig. 3. These data
clearly show that the Si adatom does not stay in its original position
which is just below the Ge atom (the Si adatom occupies a $T_4$ site just
above a second-layer Si atom on a clean surface
\cite{Takayanagi,Tong,Wang}). We conclude that Ge would prefer to
substitute the Si adatoms in its initial adsorption stages.

As shown in Figs. 2(b) and 2(c), there are three types of collective Ge 
patterns that appear on Si(111)-$7\times7$. The schematics of these Ge  
protrusions, named type-A, type-B, type-C, are given in Figs. 2(d), 2(e),
and 2(f), respectively. Type-A illustrates three Ge atoms locating at one 
corner adatom site and two adjacent center adatom sites in a HUC. Type-B
indicates the configuration with three Ge atoms occupying corner adatom   
sites in a HUC. Type-C refers to the adsorption structure with five Ge  
atoms residing on the sites of three corner adatoms and two center adatoms
in a HUC. Type-B and Type-C distribute preferentially in the FHUCs, as
shown in Figs. 2(b) and 2(c). The contrast difference between the Ge
adatoms in the type-C protrusions is attributed to both their difference
in occupation of the dangling bond states and different heights (corner
adatoms transfer less charge to the restatoms and reside higher than
center adatoms). Table I shows the site distribution of the Ge atoms. At
the coverage of 0.02 ML, the site preference ratio is about $5.6:4.4$ for
the FHUC to the UHUC, and $6.1:3.9$ for the corner to the center adatom
sites, respectively. When the coverage increases to 0.08 ML the site
preference ratios are about $9:1$ for the FHUC to the UHUC, and $4:1$ for
the corner to the center adatom sites. The site distribution for the
coverage of 0.10 ML is similar to that for the coverage of 0.08 ML. The
overall conclusion is that after an initial random occupation of Si
adatoms sites, corner adatom sites in the FHUC are preferred and gradually
type-B patterns become dominant. Type-A and Type-C patterns are more
discernible at slightly higher coverages, and, finally, small islands
begin to appear [Fig. 2(c)].

\begin{table}[tb]
\caption{Site distribution of Ge at various adatom sites at coverages of
0.02 ML, 0.08 ML, and 0.10 ML, respectively.}
\begin{ruledtabular}
\begin{tabular}{lcccccccc}
&0.02 ML&0.08 ML&0.10 ML\\
\colrule
Faulted corner sites&40\% &76\% &65\%\\
Faulted center sites&17\% &12\% &13\%\\
Unfaulted corner sites&24\%&4\% &8\%\\
Unfaulted center sites&19\% &8\%&14\%\\
\end{tabular}
\end{ruledtabular}
\end{table}

Earlier theoretical studies employing semiempirical methods suggested that
Ge atoms bond directly to Si adatoms and restatoms and reside at the
on-top sites \cite{Grodzicki,Stauffer2}. The studies also concluded that
substitution of Ge for the Si adatom was not possible \cite{Stauffer2}. On
the other hand, earlier first-principles calculations based on a
$4\times4$ supercell showed that Ge would prefer to bond at the bridge 
site ($B_2$-type) between a restatom and a first-layer Si atom
\cite{Cho1}. While these theoretical conclusions are inconsistent with our
experimental observations, we note that the possibility that Ge atoms may
replace Si adatoms was not considered in the previous first-principles
calculations \cite{Cho1}.

We performed first-principles density functional calculations using the
pseudopotential method and a plane-wave basis set \cite{Kresse}. The
Si(111) surface was modeled by repeated slabs with 4 layers of Si atoms
and 4 Si adatoms, separated by a vacuum region of 12 {\AA} (each layer
contained 16 Si atoms, corresponding to a $4\times4$ surface unit cell,
which is a small piece of the $7\times7$ cell; as in Ref. 20, this cell is
adequate for the present purposes). Two of the four restatoms were
saturated by hydrogens, so that the ratio of the number of the adatoms to
that of the restatoms is the same as for the $7\times7$ surface. Except
for the Si atoms in the bottom layer, which were fixed and saturated by H
atoms, all the atoms were relaxed until the forces on them were less than
0.05~eV/{\AA}. Exchange-correlation effects were treated with the
generalized gradient-corrected exchange-correlation functionals given by 
Perdew and Wang \cite{Perdew2}. We adopted the Vanderbilt ultrasoft
pseudopotentials \cite{Vanderbilt}. A plane-wave energy cutoff of 14.7~Ry
and the $\Gamma$ point for reciprocal space sampling were used for all the
calculations.

All the possible configurations with a Ge atom near an adatom or/and a 
restatom were calculated. Two lowest energy configurations, shown in Fig.
4, were found to have essentially the same total energy (the difference in
total energy is smaller than 0.02 eV). The first configuration consists of
Ge at a $B_2$ site [Fig. 4(a)], as identified earlier by Cho and Kaxiras
\cite{Cho1}. In the second configuration [Fig. 4(b)], the adsorbed Ge atom
substitutes for a Si adatom and the Si adatom occupies a nearby $B_2$  
site. We refer to the Ge position in the second configuration as $S_4$   
(substitutional site with four nearest-neighboring silicon atoms). The   
total energies of the configurations with Ge bonded at the on-top
positions of adatoms and restatoms are significantly higher (2.3 eV and 
1.6 eV, respectively) than the $B_2$ and $S_4$ configurations, clearly
ruling out the possibility of such configurations, which were suggested
previously on the basis of semiempirical calculations
\cite{Dev,Grodzicki,Stauffer2}. For both lowest-energy configurations
($B_2$ and $S_4$), the atom (Si or Ge) at a bridge site may diffuse within
a basin (to occupy any of the six $B_2$ sites near the restatom) and
across basins (to occupy the $B_2$ sites near different restatoms). The
diffusion barriers within a basin and across basins are about 0.5 eV (0.6
eV) and 1.0 eV (1.0 eV) for the Ge (Si) atoms, respectively, in agreement
with previous first-principles calculations \cite{Cho1,Cho2}. On the other
hand, the Ge atoms at the $S_4$ sites are not able to diffuse on their
own. Therefore, the Ge atom in an $S_4$ configuration is thermodynamically
more stable than in a $B_2$ configuration. In particular, after the atoms
initially bonded at the $B_2$ sites migrate to step edges and/or to form
islands, the surface exhibits a stable Ge-$S_4$ configuration in which Ge
atoms substitute for some of the Si adatoms and no atoms are bonded at any
of the $B_2$ sites [Fig. 4(c)], as shown by our STM observations. Small
islands, which accommodate the substituted Si adatoms, were observed in
the STM images with larger scanning areas.

It is known that the backbonds of the Si adatoms on the Si(111)-$7\times7$
surface are under considerable strain \cite{Takayanagi,Tong,Avouris}. It
is therefore expected that the adsorbed Ge atoms are able to break the
backbonds and replace the Si adatoms at elevated temperatures. Previous  
studies have established that the corner adatoms in the FHUCs are under  
more strain than the other adatoms, implying that backbonds of the corner
adatoms in the FHUCs are broken easier than those of the other adatoms
\cite{Tong,Avouris}. When Ge atoms are deposited on the surface, in
addition, the chance for the Ge atoms occupying the $B_2$ sites near a
center adatom is larger than that near a corner adatom (the center adatom
has two nearby rest atoms while the corner adatom has only one). Thus, the
Ge-$S_4$ bonding structure tends to be preferentially formed at the corner
adatom sites and in the FHUCs. Note that Ge adsorption does not result in
appreciable surface-atom relaxations, suggesting that it does not cause
strain relief.

Finally, the relaxed Ge-$S_4$ configuration obtained from our calculations
shows that the Ge atom resides at the position higher by $\sim$~0.24~{\AA}
along the direction of the surface normal than the original Si adatom that
has been replaced by Ge, in good agreement with our STM data.

In summary, the bonding structure of Ge atoms on Si(111)-$7\times7$ at low
coverages was investigated with STM and first-principles calculations. We
found that individual Ge atoms reside on the Si adatom sites and occupy
preferentially the Si corner adatom sites in the faulted half unit cells
on Si(111)-$7\times7$. STM measurements and first-principles calculations
for the geometrical structures, together with energetics from the
first-principles theory, demonstrate substitution of Ge atoms for the Si
adatoms on the Ge-adsorbed Si(111)-$7\times7$ surface.

This work was supported in part by the Natural Science Foundation of China
and Chinese National ``863" and ``973" projects, by National Science
Foundation Grant DMR-0111841, and by the William A. and Nancy F. McMinn
Endowment at Vanderbilt University. Access to the Florida State University
supercomputers is also acknowledged.

\begin{figure}[h]
\caption{(Color) Schematic top view of the Si(111)-$7\times7$
reconstruction. The outlined is the $7\times7$ unit cell with the faulted
and unfaulted half unit cells located on the left and right sides,
respectively.}
\label{autonum}
\end{figure}

\begin{figure}[h]
\caption{(Color) Filled state STM images of the Si(111)-$7\times7$ surface
with Ge coverages of (a) 0.02 ML; (b) 0.08 ML; and (c) 0.10 ML. A
$7\times7$ unit cell is marked by two triangles in (a), where F and U
represent the FHUC and UHUC, respectively. Sample bias:  $-2.2$~V in (a)
and $-1.5$~V in (b) and (c); Tunneling current: 0.5 nA in (a) and 0.2 nA
in (b) and (c). The scanning area is 20 nm $\times$ 20 nm. Three different
configurations of Ge protrusions distributions are denoted in (b) and (c)
by dot-line triangle, solid-line triangle, and dashed-line triangle,
respectively. The schematics for the three typical Ge protrusions, named
type-A (d), type-B (e), and type-C (f), are also shown.}
\label{autonum}
\end{figure}

\begin{figure}[h]
\caption{The profile lines corresponding to the dashed-arrow lines in the
STM images of Fig. 2(b) [(a)] and Fig. 2(c) [(b)], respectively.}
\label{autonum}
\end{figure}

\begin{figure}[h!]
\caption{Schematics of the minimum-energy configurations for a Ge atom on
the Si(111) surface: (a) Ge at a $B_2$ site and the nearby Si adatom at
the position off its original site; and (b) Ge at a substitutional $S_4$
site and the Si adatom at a $B_2$ site; and (c) Ge at an $S_4$ site with
the substituted Si adatom diffused away. The bond lengths are shown in
{\AA}.}
\label{autonum}
\end{figure}

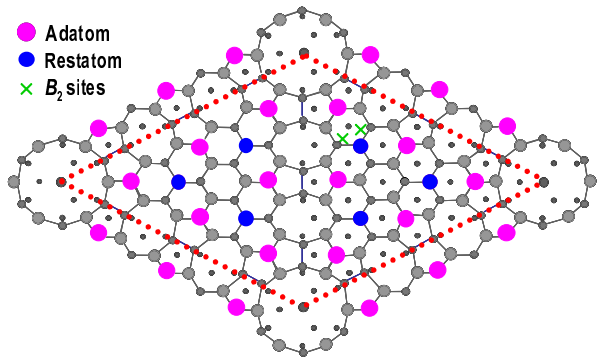
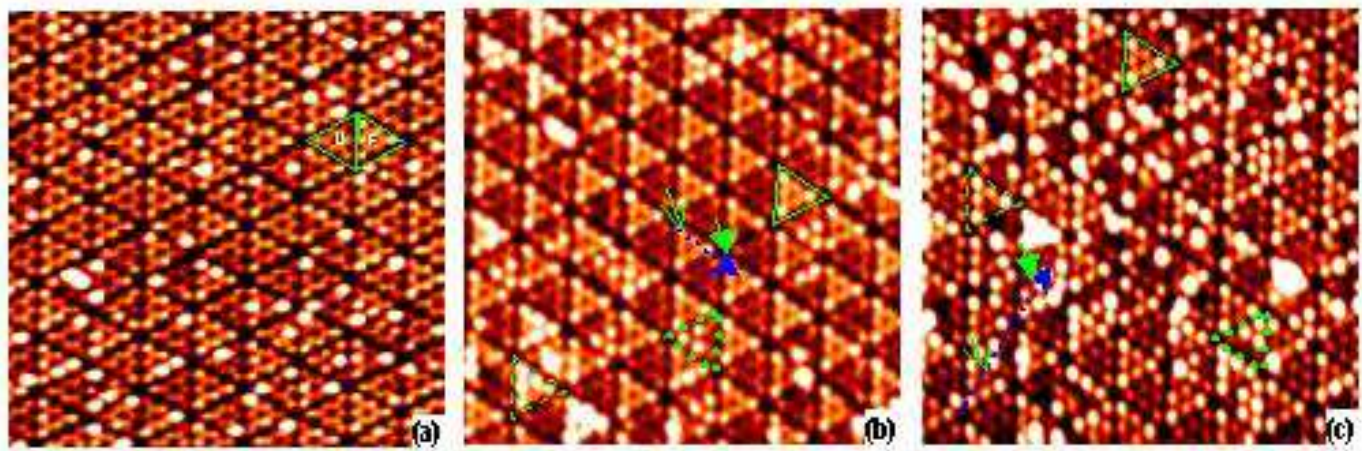
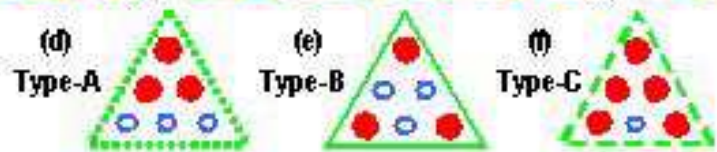
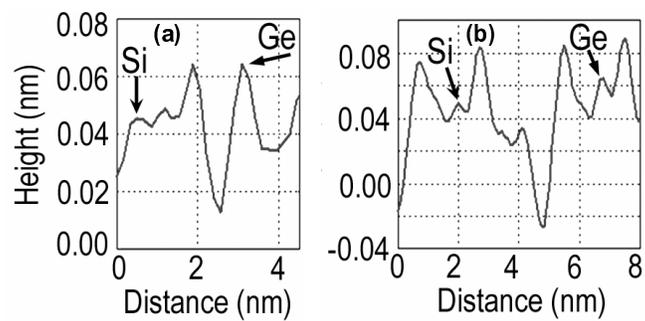
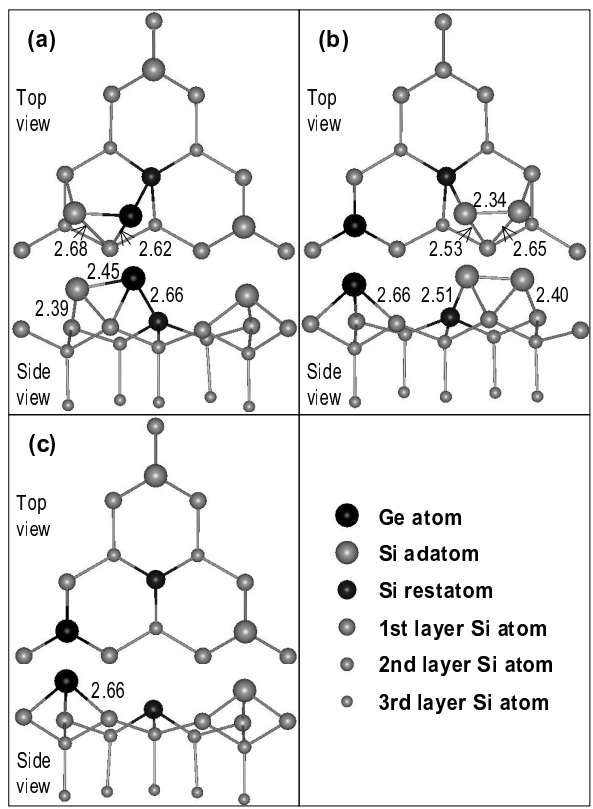
\end{document}